\documentclass[twocolumn,aps,nofootinbib]{revtex4}
\usepackage{epsfig} \usepackage{amsfonts,amssymb,amsmath}
\usepackage[colorlinks=true, pdfstartview=FitV, linkcolor=blue, citecolor=red, urlcolor=magenta]{hyperref}
\usepackage{subfigure}
\usepackage{enumerate}
\usepackage{amsmath}

\newcommand{\al}{\ensuremath{\alpha}}
\newcommand{\ga}{\ensuremath{\gamma}}

\newcommand{\ka}{\ensuremath{\kappa}}
\newcommand{\la}{\ensuremath{\lambda}}
\newcommand{\La}{\ensuremath{\Lambda}}

\newcommand{\Del}{\ensuremath{\nabla}} 
 
 \newcommand{\be}{\begin{equation}}
\newcommand{\ee}{\end{equation}} 
\newcommand{\ba}{\begin{eqnarray}} \newcommand{\ea}{\end{eqnarray}}

\newcommand{\M}{\ensuremath{\mathcal{M}}}

\newcommand{\lab}[1]{\label{#1}}
\newcommand{\bib}[1]{\bibitem{#1}}

 \def\mn{{\mu\nu}} \def\tt{\textrm}

\begin{document}

\title{Vacua in novel 4D Einstein-Gauss-Bonnet Gravity: pathology and instability?}
\author{Fu-Wen Shu$^{1,2,3}$}
\thanks{shufuwen@ncu.edu.cn}
\affiliation{
$^{1}$Department of Physics, Nanchang University, Nanchang, 330031, China\\
$^{2}$Center for Relativistic Astrophysics and High Energy Physics, Nanchang University, Nanchang 330031, China\\
$^{3}$GCAP-CASPER, Physics Department, Baylor University, Waco, TX 76798-7316, USA }

\setcounter{footnote}{0} \renewcommand{\thefootnote}{\arabic{footnote}}


\begin{abstract}
\baselineskip=0.4 cm
\begin{center}
{\bf Abstract}
\end{center}
We show an inconsistence of the novel 4D Einstein-Gauss-Bonnet gravity by considering a quantum tunneling process of vacua. Using standard semi-classical techniques,  we analytically study the vacuum decay rate for all allowed cases in the parameter space. It turns out, without exception, that the theory either encounters a disastrous divergence of vacuum decay rate, or exhibits  a confusing complex value of vacuum decay rate, or involves an instability (a large vacuum mixing). These suggest a strong possibility that the theory, at least the vacuum of the theory, is either unphysical or unstable, or has no well-defined limit as $D\rightarrow 4$.

\end{abstract}

\maketitle
\section{Introduction}
It was recognized a long time ago that the Gauss-Bonnet term in four dimensions is a topological surface term and has no contribution to the dynamical degrees of freedom of the equations of motion. The Einstein-Hilbert action together with a cosmological constant, as claimed by the Lovelock's theorem~\cite{Lanczos,Lovelock}, is the unique gravitational action giving rise to field equations which is second-order  and preserves diffeomorphism invariance. However, recent paper by Glavan and Lin \cite{Glavan:2019inb} showed a different possibility. By rescaling the coupling constant $\alpha$ of the Gauss-Bonnet term to $\alpha/(D-4)$, and defining the four-dimensional theory as the limit $D\rightarrow4$, the authors of \cite{Glavan:2019inb} propose a novel theory of gravity in 4-dimensional spacetime where the Gauss-Bonnet term gives rise to non-trivial dynamics and the theory respects Lovelock’s theorem. It is also thought to be free from the Ostrogradsky instability\cite{ostro}.  Due to these attractive features, it attracts a great deal of interests in the last few months, from black hole solutions, to properties of this novel theory, to applications in a variety of fields were explored immediately \cite{Konoplya:2020bxa, Guo:2020zmf, Casalino:2020kbt, Konoplya:2020qqh, Fernandes:2020rpa,  Lu:2020iav, Konoplya:2020ibi, Ghosh:2020syx, Konoplya:2020juj, Kobayashi:2020wqy, Zhang:2020qam, HosseiniMansoori:2020yfj, Kumar:2020uyz, Wei:2020poh, Zhang:2020sjh, Churilova:2020aca, Islam:2020xmy, Liu:2020vkh, Konoplya:2020cbv, Jin:2020emq, Heydari-Fard:2020sib,Wei:2020ght, Li:2020tlo,Kumar:2020owy}. 

A very recent work\cite{Gurses:2020ofy}, however, claimed that the novel Einstein-Gauss-Bonnet(EGB) theory in four dimensions does not have an intrinsically four-dimensional definition.  The equations of motion of the novel EGB in four dimensions are not well-defined in the limit $D\rightarrow4$. They showed this point by splitting the Gauss-Bonnet tensor into two parts: the part has an explicit ($D-4$) factor in front of it, and the part which does not have an explicit ($D-4$)  factor which is called the Lanczos-Bach tensor and vanishes identically in four dimensions. The subtlety is that the Lanczos-Bach tensor cannot be identically zero since if it does so, the Gauss-Bonnet tensor does not satisfy the Bianchi Identity and thus the whole equations of motion do not hold.  Accordingly, the solutions of the theory are defined in $D>4$ dimensions and the Lanczos-Bach tensor has no well-defined limit as $D\rightarrow4$.  In addition, very recently some other issues \cite{Ai:2020peo,Malafarina:2020pvl,Mahapatra:2020rds,1792021}were raised, making the validity of the theory questionable.

In this work we provide another evidence of the pathology of the theory.  We focus on properties, especially the stability, of the vacua of the theory. Our efforts are divided in two folds: Classically, we find that the pure vacuum itself is free of the subtlety found in \cite{Gurses:2020ofy}, and is stable at the classical level though it has a ghost coupling for the Gauss-Bonnet vacuum. However, once it couples with matter, non-spherically symmetric solutions can excite a freely propagating tensor mode  \cite{Charmousis:2008ce}. As a consequence, the ghost coupling gives rise to negative energy gravitational waves, causing the Gauss-Bonnet background unstable.  More serious problem occurs at quantum level. We find a pathology in the vacuum tunneling process.  With the help of standard semi-classical techniques\cite{semiclassical1,semiclassical2}, we analytically study the vacuum decay rate for all allowed values of parameters. Our results show that the theory either suffers a divergent vacuum decay rate, or possesses a puzzling complex value of the decay rate, or appears a large vacuum mixing. All these strongly indicate that the theory, at least the vacuum of the theory, is either unphysical or unstable, or has no well-defined limit as $D\rightarrow 4$.

\section{Vacua of the theory}
Following the logic of \cite{Glavan:2019inb}, we first keep everything in arbitrary $D$ dimensions, and in the end we take the limit $D\rightarrow4$. The novel Einstein-Gauss-Bonnet (EGB) gravity in $D$ dimensions is described as a limit of the following action
\be \lab{gbaction}
S_{EGB}=\kappa \int_\M d^D x \sqrt{-g}(R-2\Lambda_0+\frac{\al}{D-4} L_{GB} ),
\ee
where $L_{GB}=R^2-4R_{ab}R^{ab}+R_{abcd}R^{abcd}$ is the Gauss-Bonnet term, $\al$ is the coupling constant and $\Lambda_0$ is the bare cosmological constant.  Note that in writing \eqref{gbaction} we have rescaled the coupling constant by $\alpha\rightarrow \frac{\al}{D-4}$.
 The corresponding field equations are given by
\be\label{eom}
    R_{ab}-\frac12g_{ab}R+\La_0 g_{ab} +\frac{\al}{D-4} H_{ab}=0,
\ee
where $H_{ab}$ is the Gauss-Bonnet tensor which is of the form
\ba
H_{ab}&=&2\left(RR_{ab}-2R_{ac}R_b^c-2R_{acbd}R^{cd}+R_{acde}R_b{}^{cde}\right.\nonumber\\
&&\left.-\frac14g_{ab} L_{GB} \right).
\ea
 The theory admits two maximally symmetric vacuum solutions, with two possible effective cosmological constants,
\be
\Lambda^\pm_\tt{eff}=\Lambda_{CS}\left(1 \pm \sqrt{1-\frac{2 \Lambda_0}{\Lambda_{CS}}}\right), \lab{Lameff}
\ee
where $ \Lambda_{CS}$ is known as the  Chern-Simons limit
\be
 \Lambda_{CS}=-\frac{1}{4\alpha} \frac{(D-1)(D-2)}{(D-3)}.
\ee
For the vacua to be well defined, we have $\La_0\leq\La_{CS}/2$. Generally, two solutions are distinct except for $\La_0=\La_{CS}/2$ where two vacua degenerate. The  lower root $\La_\tt{eff}^- $ is often referred to as the Einstein branch as it goes smoothly to the Einstein vacuum, $\La_\tt{eff}^- \to \La_0$, as $\alpha \to 0$. The other one, $\La_\tt{eff}^+$, instead, has a limit $\La_\tt{eff}^+ \to 2\La_{CS}\sim -1/\alpha$ as $\al \to 0$, which has no related counterpart in Einstein's gravity.  It is a result of Gauss-Bonnet corrections and thus it is usually known as the ``stringy'' or ``Gauss-Bonnet'' branch.

Note that the above vacuum solutions \eqref{Lameff}  are free of the aforementioned subtleties as observed in \cite{Gurses:2020ofy}. To see this explicitly, as suggested in \cite{Gurses:2020ofy} we split the Gauss-Bonnet tensor into two parts: $H_{ab}=2(L_{ab}+Z_{ab})$, where $L_{ab}$ is the Lanczos-Bach tensor which has no explicit coefficient related to the number of dimensions. It is easy to check that the Lanczos-Bach tensor is identically vanishing for any dimension in this case.  While $Z_{ab}$ carries explicit coefficients regarding the number of dimensions which is of the following form for the above vacuum solutions
\be
Z_{ab}=-\frac{(D-4)\La_{eff}^2}{(D-1)(D-2)}g_{ab}.
\ee
As a consequence
\be
S_{ab}:=\lim_{D\rightarrow4}\left(\frac{2\alpha}{D-4}\right)Z_{ab}= -\frac{\alpha\La_{eff}^2}{3}g_{ab},
\ee
which shows that the Bianchi identity $\Del^{a}S_{ab}=0$ always hold. 

Another issue is concerning stability of the vacua. The stability at the classical level of the vacua in EGB in $D$($>4$) dimensions has been discussed widely \cite{des, gbbhs, yang, tekin, Charmousis:2008ce}. It is generally thought that the Einstein branch is classically stable while the Gauss-Bonnet vacuum contains a ghost due to the wrong sign of the perturbative action. This argument can be naively generalized to the present case. To see this clearly, let us perturb the metric around the vacua, $g_{\mu\nu}=\bar{g}_{\mu\nu}+h_{\mu\nu}$ with $\bar{g}_{\mu\nu}$ the metric of the vacua. We then expand (\ref{gbaction}) and obtain the following quadratic action
\be
S^{(2)}=-\frac{1}{2} \ka  \La_{g} \int_\M d^D x \sqrt{-\bar g}~ h^{ab} \left(G_{ab}^{L}+\La_\tt{eff} h_{ab}\right)
\ee
where $G_{ab}^{L}$ is the linearized Einstein tensor and $\La_g=\left(1-\La_\tt{eff}/\La_{CS}\right)$. The corresponding linearized field equations are
\be
 \La_g\left(G_{ab}^{L}+\La_\tt{eff} h_{ab}\right)=0. \lab{lineom}
\ee
For the Einstein branch we always have $\La_g >0 $ regardless of $\alpha$ so it is well behaved. In contrast, it has $\La_g <0 $ on the Gauss-Bonnet branch, so the kinetic term of the gravitational action has the wrong sign, indicating the presence of a ghost.  Note that presence of a ghost does not necessarily imply perturbative instabilities since a ghost can only render instability if it is freely propagating. The diffeomorphism invariance of the present EGB theory implies that the only freely propagating mode of the theory is the tensor mode. Since the vacuum (spherically symmetric) solutions only excite scalars, which are not freely propagating, the pure vacuum itself is stable at the classical level. However, as observed in \cite{Charmousis:2008ce} that, if coupled matter is considered, there is the possibility that non-spherically symmetric solutions exist (a binary system, for example) and excite the freely propagating tensor modes (gravitational waves). In this case, the ghost coupling will give rise to negative energy gravitational waves, causing the Gauss-Bonnet background unstable.  In next section, however, we will show that at the quantum level, even for pure vacuum, it encounters instability.

\section{Vacuum decay of the theory}

In this section, we show that even in pure vacuum, the theory exhibits a quantum instability. Even worse, we find there is an unacceptable divergence of the vacuum decay rate as we take the limit $D\rightarrow 4$.  We show this by investigating the tunneling rate between two vacua, through the well-known bubble nucleation process\cite{semiclassical1,semiclassical2}. It says that a quantum tunneling process from a false vacuum to a true one is realized by the nucleation of a true vacuum bubble which is surrounded in a false vacuum. 

In semiclassical approximation, the decay rate is given by $\Gamma \sim A e^{-B/\hbar}$, where the factor $A$ has been discussed in \cite{semiclassical2} and the exponent $B$ is the difference of Euclidean actions between the true vacuum solution(the bounce) and false vacuum solution(the background), namely,
$
B=S_\tt{true}-S_\tt{false}.
$
\subsection{Bubble geometry}
Let us start with the bubble geometry of the theory.  In the absence of gravity, the geometry is always Minkowski and  any local minimum of potential for a scalar field can be viewed as a vacuum of the scalar field. Energy density of the field at the vacuum can be positive, zero or negative.  The true vacuum is referred to as the vacuum with the lowest energy density. In contrast, all others are called false vacua.  When gravity is taken into consideration, the vacua are those with maximally symmetric spaces, namely,  de Sitter, Minkowski and anti-de Sitter spacetimes, respectively.  As to the EGB theory, there are generally two vacua, the one with $\La_\tt{eff}^+$ (the Gauss-Bonnet vacuum) and the one with $\La_\tt{eff}^-$ (the Einstein vacuum).  In the thin wall approximation, the two vacua are separated by a domain wall, composed of ordinary matter.

To proceed, let us follow \cite{Charmousis:2008ce} and assume that $\M_1$ be the interior of the bubble and $\M_2$ be the exterior, both of them are the maximally symmetric vacuum solution to the  field equations \eqref{eom} \footnote{Without loss of the generality, throughout the paper we assume that the bare cosmological constant $\La_0$ and $\alpha$ are the same for  $\M_1$ and  $\M_2$.}. The wall is the common boundary of the two manifolds, that is, $\Sigma=\partial\M_1=\partial\M_2$. By introducing a Gaussian coordinates $x^a=(\xi,x^{\mu})$ and supposing that the domain wall is located at $\xi=\xi_0$,  the bulk metric then can be cast to the form
\be
ds^2=g_{ab}dx^a dx^b=d\xi^2+a(\xi)^2ds_{wall}^2, \lab{bulkg}
\ee
where $a(\xi)=\rho(\xi)/{\rho_0}$ is a dimensionless scale factor and $ds_{wall}^2=\rho_0^2(-d\tau^2+\cosh^2\tau d \Omega_{D-2})$ denotes the metric of the wall, which is a de Sitter hyperboloid of constant radius $\rho_0(=\rho(\xi_0))$ embedded in the bulk geometry.  $\rho(\xi)$ is of the following form
\begin{equation}    \rho(\xi)=
 \begin{cases}
    \frac{1}{k_1}\sinh(k_1\xi), &  0\leq \xi\leq \xi_0 \lab{rho1} \\
  \frac{1}{k_2}\sinh[k_2(\xi-\beta)], \qquad &\xi_0 \leq \xi \leq \xi_\tt{max} \ \ 
 \end{cases}                \end{equation}
where 
$
k_i^2=-\frac{2\La_\tt{eff}^{(i)}}{(D-1)(D-2)} \label{ki}
$ and
\be
\xi_\tt{max}=\begin{cases}\infty & \tt{for $k_2^2 \geq 0$} ,\\
\beta +\pi/|k_2| & \tt{for $k_2^2<0$}.\end{cases} \lab{ximax}
\ee
Several boundary conditions at the wall should be imposed. On one hand, the continuity of the induced metric\eqref{bulkg} is required
\be \lab{cont}
\Delta\left[ \frac{\sinh k\la}{k}\right]=0,
\ee
where $\Delta X=X_2-X_1$ is the jump operator and $\la_1=\xi_0,\la_2=\xi_0-\beta$ respectively.

On the other hand,   on the wall the following junction conditions are needed~\cite{gbjunc}
\be
\ka \Delta \left[K_\mn-K \ga_\mn+2\al \left(Q_\mn-\frac{1}{3} Q \ga_\mn \right)\right]=\frac{\sigma}{2}\gamma_\mn, \lab{junc}
\ee
 where $\sigma$ is the tension of the wall and $K_\mn$ is the extrinsic curvature of the wall. $Q_{\mu\nu}$ is a mixed term of extrinsic curvature and the Riemann curvature of the wall, whose explicit form is given by Eq. (18) of \cite{Charmousis:2008ce}. After using \eqref{cont}, the junction condition \eqref{junc} leads to the following expression for the wall tension \cite{Charmousis:2008ce}
\be
\sigma=-\frac{2}{3}(D-2)(D-3) \ka \al \left(\frac{\Delta \left[\cosh k\la \right]}{\rho_0}\right)^3 \lab{sigma2}.
\ee
The energy momentum tensor, $T_{ab}=-\delta (\xi-\xi_0)\sigma \ga_\mn \delta^\mu_a \delta^\nu_b$ on the wall must satisfy the weak energy condition, if we require the wall is made up of ordinary fields. This is equivalent to require the tension to be nonnegative, which is satisfied if, and only if, one of the following holds \cite{Charmousis:2008ce}:
 (i)$\al>0, ~\La^\tt{eff}_1 \leq \La_2^\tt{eff} \leq 0$;
(ii)$\al>0, ~\La_1^\tt{eff} \leq 0 \leq \La_2^\tt{eff}$;
(iii)$\al<0, ~0\leq \La_1^\tt{eff} \leq \La_2^\tt{eff}, ~|k_1|\la_1 \geq |k_2| \la_2  \geq   \frac{\pi}{2}$;
(iv)$\al<0, ~\La_2^\tt{eff} \leq 0 \leq \La_1^\tt{eff}$;
(v) $\al<0, ~0\leq \La_2^\tt{eff} \leq \La_1^\tt{eff}, ~ |k_2|\la_2 \leq |k_1| \la_1 \leq \frac{\pi}{2}$.
The first three cases describe decay of the false vacuum to true vacuum. The last two cases, however, describe the reverse process, in which true vacuum decay occurs.
 
\subsection{Bubble nucleation} 
Using the standard semiclassic techniques, in appendix \ref{aa} we show that the exponent $B$ in the present case is given by
\be
B=\frac{\kappa\Omega_{D-1}}{D-4}\Theta(D,k\lambda), \label{B}
\ee
where $\Theta(D,k\lambda)$ is given by \eqref{theta}.
In what follows we will discuss in detail the vacuum decay for all five cases mentioned in the last section, case by case. 

\subsubsection{Case (i): $\al>0, ~\La^\tt{eff}_1 \leq \La_2^\tt{eff} \leq 0$}
For this case, we have $\La_{CS}<0$, $\La^\tt{eff}_1=\La^{+}_\tt{eff}, \La^\tt{eff}_2=\La^{-}_\tt{eff},$ $k_1\geq k_2\geq0$,  and 
\be
-\frac{(D-1)(D-2)}{8\al (D-3)}\leq\La_0\leq 0.\label{la01}
\ee

Substituting \eqref{f1} into \eqref{theta} and taking $D\rightarrow 4$ limit, after lengthy algebraic calculations,  we finally obtain

\ba
\Theta(D\rightarrow 4,k\la)&=&-3\al \sqrt{2(\gamma^2-\delta^2)}\left(\sqrt{\gamma+\delta}-\sqrt{\gamma-\delta}\right),\nonumber\\
\label{theta1}
\ea
where 
\be
\gamma=2+\frac{\rho_0^2}{\al},\ \ \delta=\sqrt{1+\frac{4\al\La_0}3} \frac{\rho_0^2}{\al},
\ee
and we have used $k_i\la_i=\sinh^{-1}(k_i\rho_0)$.
In the limit $D\rightarrow 4$, \eqref{la01} becomes $-\frac{3}{4\al}\leq\La_0\leq 0$, which implies $0\leq\delta\leq \frac{\rho_0^2}{\al}$ and $\gamma>\delta$ thus $\gamma+\delta\geq\gamma-\delta>0$. Therefore $\Theta$ is always negative except when $\La_0=-3/(4\al)$, where $\Theta=0$.  

Back to the decay rate \eqref{B}, this introduces a negative divergence as $D\rightarrow 4$ since $\Theta$ is negative nonzero finite value except exactly at  $\La_0=-3/(4\al)$. Translating into $\Gamma \sim A e^{-B/\hbar}$ it leads to a catastrophic infinity of the vacuum decay rate which is not well defined and is definitely unphysical.  Any infinitesimal deviation of $\La_0=-3/(4\al)$ will suffer this pathology, which strongly suggests the theory is not well defined as $D\rightarrow4$. Even at $\La_0=-3/(4\al)$, where $B=0$ and $\Gamma\sim 1$, there is a large mixing between two distinct vacua, implying an instability in the sense that the rate of particle production diverges, and the vacuum is destroyed infinitely quickly \cite{tanaka}.

\subsubsection{Case (ii): $\al>0, ~\La_1^\tt{eff} \leq 0 \leq \La_2^\tt{eff}$}
Now we have $\La_{CS}<0$, $\La^\tt{eff}_1=\La^{+}_\tt{eff}, \La^\tt{eff}_2=\La^{-}_\tt{eff},$ $k_1\geq 0$, while $k_2$ is imaginary, and 
$\La_0\geq 0$.

Again by substituting \eqref{f1} and  \eqref{f2} into \eqref{theta}, using $k_1\la_1=\sinh^{-1}(k_1\rho_0), k_2\la_2=i\sin^{-1}(|k_2|\rho_0)$, and taking $D\rightarrow 4$ limit, finally we obtain the same result as \eqref{theta1}.

In this case $\gamma+\delta>\gamma-\delta$ always holds. As $0\leq\La_0\leq 3(\al+\rho_0^2)\rho^{-4}$, we have $\gamma\geq \delta$, then as before, we find $\Theta\leq0$, implying a divergence of $B$ as $D\rightarrow 4$. For $\La_0>3(\al+\rho_0^2)\rho^{-4} $, one gets $\gamma < \delta$, implying $\Theta$ is complex. So far we do not know if it has any physical meaning for analytical continuation  to complex domain. Even it has, one can easily show  that the real part of $\Theta$ is always negative. This again leads to a catastrophic divergence of $B$ as $D\rightarrow 4$.

\subsubsection{Case (iii): $\al<0, ~0\leq \La_1^\tt{eff} \leq \La_2^\tt{eff}, ~|k_1|\la_1 \geq |k_2| \la_2  \geq   \frac{\pi}{2}$}
In this case we have $\La_{CS}>0$, $\La^\tt{eff}_1=\La^{-}_\tt{eff}, \La^\tt{eff}_2=\La^{+}_\tt{eff},$ $k_1$ and $k_2$ are imaginary with $|k_1|\leq |k_2|$, and 
$0\leq\La_0\leq -\frac{(D-1)(D-2)}{8\al (D-3)}$ ($0\leq\La_0\leq-\frac3{4\al}$ as $D\rightarrow 4$).

Following the same procedures as above cases, we substitute \eqref{f2} into \eqref{theta} and note that $k_i\la_i=\pi i-i\sin^{-1}(|k_i|\rho_0)$  such that $|k_1|\la_1 \geq |k_2| \la_2  \geq   \frac{\pi}{2}$ can be fulfilled. The same result\eqref{theta1}, once again, is obtained in the limit $D\rightarrow 4$. 

By definition we find $\frac{\rho_0^2}{\al}\leq\delta\leq0$ and $\gamma+\delta\leq\gamma-\delta$. Therefore, there are four possibilities of the value of $\Theta$: (a) As $\gamma+\delta<0$ and $\gamma-\delta>0$, $\Theta$ is complex and its real part is negative; (b)  As $\gamma+\delta>0$, $\Theta$ is real and negative; (c) As $\gamma-\delta<0$, $\Theta$ is pure imaginary; (d) For $\delta=0$ or $\gamma=\pm\delta$, then $\Theta=0$.  In all cases, there is either a complex vacuum decay rate that is physically incomprehensible, or there is divergence of $B$ as $D\rightarrow 4$, or there is a large vacuum mixing. Hence the theory is either unphysical or unstable.

\subsubsection{Case (iv): $\al<0, ~\La_2^\tt{eff} \leq 0 \leq \La_1^\tt{eff}$}
This case corresponds to $\La_{CS}>0$, $\La^\tt{eff}_1=\La^{+}_\tt{eff}, \La^\tt{eff}_2=\La^{-}_\tt{eff},$ $k_2\geq 0$, while $k_1$ is imaginary, and 
$\La_0\leq 0$.  It is straightforward, following the procedures addressed above but with $k_1\la_1=i\sin^{-1}(|k_1|\rho_0)$ and $k_2\la_2=\sinh^{-1}(k_2\rho_0)$, to show that $\Theta$ has the same form as \eqref{theta1} after taking the $D\rightarrow 4$ limit. 

Now we have $\gamma+\delta<\gamma-\delta$ and $\gamma-\delta\geq 2$. When $\gamma+\delta\geq 0$(corresponds to $3(\al+\rho_0^2)\rho^{-4}\leq\La_0\leq0\ \&\ \al+\rho_0^2<0$ for $\Theta<0$ and $\al+\rho_0^2=0\ \&\ \La_0=0$ for $\Theta=0$),  we have $\Theta\leq0$, indicating a divergence of $B$ as $D\rightarrow 4$. While as $\gamma+\delta< 0$(corresponds to $\La_0\leq 3(\al+\rho_0^2)\rho^{-4}\ \&\ \al+\rho_0^2<0$, or $-\rho_0^2<\al<0\ \&\ \La_0=0$),  $\Theta$ is complex with $Re(\Theta)<0$. Whatever it means in physics, this again leads to a catastrophic divergence of $B$ as $D\rightarrow 4$.

\subsubsection{Case (v): $\al<0, ~0\leq \La_2^\tt{eff} \leq \La_1^\tt{eff}, ~ |k_2|\la_2 \leq |k_1| \la_1 \leq \frac{\pi}{2}$}
Variables of this case satisfy: $\La_{CS}>0$, $\La^\tt{eff}_1=\La^{+}_\tt{eff}, \La^\tt{eff}_2=\La^{-}_\tt{eff}$, both $k_1$ and $k_2$ are imaginary with $|k_2|\leq |k_1|$, and 
$0\leq\La_0\leq -\frac{(D-1)(D-2)}{8\al (D-3)}$ ($0\leq\La_0\leq-\frac3{4\al}$ as $D\rightarrow 4$). As expected, the same result \eqref{theta1} is obtained after the standard aforementioned procedures, except that $k_i\la_i=i\sin^{-1}(|k_i|\rho_0)$  such that $|k_2|\la_2 \leq |k_1| \la_1 \leq \frac{\pi}{2}$ is preserved. 

It is easy to check that we have all the same four cases as discussed in the case (iii). Therefore,  all the statements made there are also reliable here, that is, the theory is either unphysical or unstable.

\begin{figure}
\centering
\includegraphics[scale=0.42]{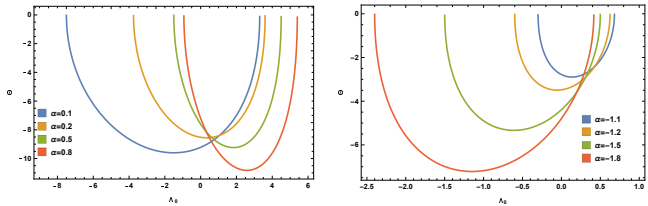}
\caption{The vacuum decay function $\Theta$ as a function of $\La_0$ for different values of $\al$ with fixed $\rho_0=1$. The left panel refers to as cases with positive $\al$, while the right panel corresponds to the negative one. It shows that we cannot find positive value for $\Theta$ in the parameter space.} \label{fig1}
\end{figure}

In summary, none of all these cases gives a positive finite value of $\Theta$. It is either real nonpositive or complex with nonpositive real part. In order to have a numeric confirmation, in  FIG.\ref{fig1} we plot $\Theta$ as a function of $\La_0$ for different values of $\al$ and fixed $\rho_0$. We show the numeric result agrees with our analytic analysis.

\section{Summary}
In this work we find a pathology of a recent proposed novel $D=4$ Einstein-Gauss-Bonnet gravity. By investigating vacuum bubble nucleation process, we find that, although the vacuum itself is free of the subtlety addressed in \cite{Gurses:2020ofy} and is stable at the classical level, it suffers from a quantum pathology. Specifically, we study in detail all cases allowed for bubble nucleation. We find,  without exception, that the theory either suffers a catastrophic divergence of the vacuum decay rate, or has a complex vacuum decay rate that is physically incomprehensible, or possesses a large vacuum mixing. All these force us to conclude that the theory, at least the vacuum of the theory, is either unphysical or unstable, or has no well-defined limit as $D\rightarrow 4$.

Our analysis is essentially different from the one given in \cite{Charmousis:2008ce} at least in three aspects: First, we do not need to restrict $\La_\tt{eff}^{(i)}$ close to the Chern-Simons limit $\La_{CS}$, nor need to make a perturbative expansion around $\La_{CS}$. Our result, without need of perturbative expansion, is reliable to all possible values of $\La_\tt{eff}^{(i)}$ allowed by the standard bubble nucleation computation \cite{semiclassical2}; Second, our results show an intrinsic pathology of the decay rate of the theory. Their result does not have any divergence concerning $B$, and the theory possibly stable when appropriate UV cut-off is introduced. Our case, however, $B$ itself is divergent everywhere in the whole parameter space except two points where $B=0$, indicating the theory itself is not well defined in the limit  $D\rightarrow 4$; The last point, more straightforwardly, is that the present case focus on the novel $4D$ EGB model, while their work pay attention to the ordinary $D$ ($D\geq 5$) dimensional EGB gravity.

\section*{Acknowledgements}

This work was supported in part by the National Natural Science Foundation of China under grant numbers 11975116, 11665016, and Jiangxi Science Foundation for Distinguished Young Scientists under grant number 20192BCB23007.
\\
\ \\
{\bf Note added}: After this work was completed, we learned a work \cite{Fernandes:2020nbq} , which appeared in arXiv a couple of days before. They proposed a regularization scheme to cancel the divergence. The other paper\cite{Hennigar:2020lsl}, which appeared in arXiv in the same day, proposed a similar approach. It is an interesting future work to see if it still works for our present issue.
\appendix
\section{Vacuum decay rate}\label{aa}
Following the procedures given in \cite{Charmousis:2008ce} , it is not difficult to find that $B$ in the present case is given by
\be
B=\frac{\kappa\Omega_{D-1}}{D-4}\Theta(D,k\lambda), \label{A1}
\ee
where
\ba
\Theta(D,z) &=& 
(4-D)\Big(2\La_0 \Delta [F(z)]+D(D-1)\Delta[k^2F(z)] \nonumber\\ 
&&-\frac{D\cdots(D-3) \al}{D-4}\Delta[ k^4 F(z)]\Big)+\nonumber\\ 
&&2\rho_0^{D-1}\left(-2\left(D-1\right)\frac{\Delta \left[\cosh z\right]}{\rho_0}\right.\nonumber\\ 
&&\left.+(D-2)(D-3)\al \left(\frac{\Delta \left[\cosh z\right]}{\rho_0}\right)^3\right).\lab{theta}
\ea
where $F(z)$ is defined as
\be
F(z)=k^{-D}\int_0^z dy \left(\sinh y\right)^{D-1}.
\ee
It is straightforward to show that the integral of $F(z)$ can be integrated out to obtain
\ba
F(z)&=&\frac{i k^{-D}\pi ^{3/2}}{\left(e^{i \pi  D}-1\right) \Gamma \left(1-\frac{D}{2}\right) \Gamma \left(\frac{D+1}{2}\right)}\nonumber\\ 
&&-(ik)^{-D} \cosh (z) \, _2F_1\left(\frac{1}{2},1-\frac{D}{2};\frac{3}{2};\cosh ^2(z)\right),\nonumber\\ \label{f1}
\ea
for $z$ is real and $z\geq0$, and
\ba
F(z)&=&\frac{\pi ^{3/2}  \left(ik\right)^{-D} \csc \left(\frac{\pi  D}{2}\right)}{2 \Gamma \left(1-\frac{D}{2}\right) \Gamma \left(\frac{D+1}{2}\right)}\nonumber\\ 
&&-\left(ik\right)^{-D}\cosh (z)  \, _2F_1\left(\frac{1}{2},1-\frac{D}{2};\frac{3}{2};\cosh ^2(z)\right),\nonumber\\ \label{f2}
\ea
for $z$ is imaginary.

\end{document}